# RISK-BASED TEST FRAMEWORK FOR LLM FEATURES IN REGULATED SOFTWARE

Zhiyin Zhou

New York, USA

*ABSTRACT*

*Large language models are increasingly embedded in regulated and safety critical software, including clinical research platforms and healthcare information systems. While these features enable natural language search, summarization, and configuration assistance, they introduce risks such as hallucinations, harmful or out of scope advice, privacy and security issues, bias, instability under change, and adversarial misuse. Prior work on machine learning testing and AI assurance offers useful concepts but limited guidance for interactive, product embedded assistants. This paper proposes a risk-based testing framework for LLM features in regulated software: a six-category risk taxonomy, a layered test strategy mapping risks to concrete tests across guardrail, orchestration, and system layers, and a case study applying the approach to a Knowledgebase assistant in a clinical research platform.*

## 1. INTRODUCTION

Large language models (LLMs) have rapidly become core components of modern software systems, enabling natural language search, summarization, decision support, and conversational interfaces across many domains [4], [8]. In regulated settings such as healthcare, finance, and clinical research, these capabilities are increasingly embedded directly into workflows for clinicians, coordinators, and operations staff [20]–[22], [26]. While LLMs promise efficiency and improved usability, they also introduce new classes of failure modes. These failures are not only technical defects but can manifest as safety incidents, privacy breaches, or regulatory non-compliance when they occur in high-stakes contexts [3], [21], [26].

A growing body of work has documented that LLMs can produce fluent yet factually incorrect content, a phenomenon often referred to as hallucination [4], [5], [7]. Benchmarks such as TruthfulQA and HELM show that even state of the art models struggle with factual consistency, especially on knowledge-intensive or domain specific queries [4], [5]. At the same time, research on alignment and safety has shown that models may generate harmful, biased, or otherwise inappropriate content unless carefully trained and constrained [8]–[10], [25], [30]. These risks are amplified in domains where outputs may influence clinical decisions, protocol configuration, patient communication, or financial actions [16], [20], [21], [26].

Software engineering and machine learning communities have proposed numerous techniques for testing and validating ML systems more broadly [1]–[3]. Surveys and safety frameworks highlight challenges such as non-determinism, data dependence, and unclear test oracles, and advocate for combinations of model-centric validation, simulations, expert review, and post-deployment monitoring [1]–[3], [17]–[19]. In safety-critical sectors, assurance frameworks like AMLAS and domain specific guidance from regulators further emphasize the need for structured safety arguments and evidence spanning the entire system lifecycle [17], [18], [20]–[22], [24]. However, existing work primarily targets classifiers and perception models rather than interactive, generative LLM features. It provides limited guidance on how to derive test plans from LLM-specific risks such as jailbreak prompts, prompt injection, unstable behavior under model updates, and content policy violations [11]–[15].

In parallel, regulators and international bodies have begun to articulate expectations for trustworthy AI. FDA guidance for AI-enabled medical devices, the WHO framework for AI in health, and emerging AI risk management standards all stress robustness, transparency, bias control, and ongoing performance monitoring [20], [21], [24]–[26]. Yet these documents are largely technology agnostic; they describe what must be demonstrated but not how teams should concretely test and monitor LLM-based features embedded in software products. There is therefore a gap between high level regulatory principles and the day-to-day testing practices of engineering teams building LLM-enabled functionality.

This paper addresses that gap by proposing a risk-based testing framework for LLM features in regulated software systems. We make three contributions. First, we synthesize prior work on LLM evaluation, safety, and AI assurance into a taxonomy of risk categories that is tailored to regulated environments, covering factual errors, harmful or out-of-scope advice, privacy and security leakage, bias and unfairness, instability under change, and adversarial misuse [3]–[5], [7]–[12], [16], [21], [25], [26], [30]. Second, we derive a structured test strategy that maps each risk category to specific test types and automation points, integrating traditional ML testing ideas with LLM-oriented techniques such as red teaming and prompt-based regression suites [1], [2], [11]–[15], [17], [19]. Third, we discuss how this strategy can be aligned with validation practices and governance expectations in healthcare and related regulated domains, outlining how the resulting test artefacts support safety arguments and audits [20]–[22], [24]–[26]. Together, these elements aim to provide practitioners with a pragmatic starting point for systematically testing LLM-based features when product failures have consequences beyond simple user dissatisfaction.

## 1.1 Approach

The work presented in this paper is based on a combination of narrative literature synthesis and design-oriented method development grounded in a concrete product scenario. Rather than conducting a formal systematic review, we focused on integrating insights from four strands of prior work that are directly relevant to testing LLM based features in regulated settings.

First, we surveyed surveys and mapping studies on testing and validation of machine learning systems, including model centric testing techniques, lifecycle perspectives, and safety assurance frameworks for learning enabled components [1]–[3], [17]–[19]. These sources provided concepts such as test oracles for non-deterministic models, continuous validation, and structured safety arguments that informed the architectural decomposition and the emphasis on regression and monitoring.

Second, we examined research on evaluation and safety of large language models, including holistic benchmarks, factuality tests, hallucination detection methods, alignment techniques, and red teaming practices [4]–[15]. From this body of work, we extracted recurring concerns about factual inconsistency, harmful content, adversarial prompts, and the limits of training time alignment as the sole safety measure.

Third, we reviewed literature and guidance on validation and governance of AI systems in healthcare and related regulated domains, including analyses of AI enabled medical devices, clinical validation approaches, and ethics and governance frameworks [3], [20]–[26], [28]. These texts clarified the expectations of regulators and professional bodies regarding evidence, documentation, bias monitoring, and lifecycle management.

Fourth, the design of the test strategy was informed by practical experience with integrating and testing LLM based knowledgebase functionality in a clinical research software platform. This experience provided concrete examples of failure modes and test artefacts, which are reflected in the illustrative case in Section 4.4.

The risk taxonomy in Section 3 was derived by clustering the concerns identified in the LLM safety and regulated AI literature into categories that are meaningful for software engineering teams responsible for embedded features. The test types in Section 4 were then constructed by mapping these categories onto known machine learning testing techniques and safety assurance practices, and by refining them through their application to the knowledgebase assistant. The resulting framework is thus a synthesis of existing research, regulatory expectations, and product level experience, rather than a purely theoretical construction.

## 2. BACKGROUND

This section summarizes prior work that informs how LLM features should be tested when they are embedded inside regulated products. We focus on (1) testing and assurance for machine learning systems and (2) evaluation and safety research for LLMs.

### 2.1. Testing and validation of machine learning systems

Traditional software testing relies on deterministic behavior and well-defined test oracles. Machine learning systems violate both assumptions. The same input may yield different outputs due to stochastic inference or retraining, and correct behavior is often defined statistically rather than as a single expected value. Recent surveys on machine learning testing highlight these challenges and propose new taxonomies of faults, test objectives, and test generation techniques [1]–[3]. Common themes include the need to test not only model code but also data pipelines, feature engineering, and model integration into larger software systems.

Zhang et al. categorize machine learning specific faults such as label noise, data drift, and adversarial examples, and discuss approaches like metamorphic testing, differential testing, and data augmentation-based test generation to reveal them [1]. Riccio et al. provide a mapping study that shows how testing activities span the entire machine learning lifecycle from data collection to deployment, and that many proposed techniques focus on classification models with structured inputs [2]. Myllyaho et al. extend the discussion to validation at system level, identifying four broad families of validation methods for AI systems: model centered evaluation on curated datasets, simulation-based validation, prospective trials, and expert based assessment [3]. These works motivate a view of AI validation as a combination of offline evaluation, scenario based testing, and human review rather than a single pass test phase.

In safety critical domains such as autonomous driving, the software assurance community has begun to translate these ideas into structured safety arguments and processes. Frameworks like AMLAS define how to construct a safety case for machine learning components, including hazard analysis, data management plans, and evidence of robustness under relevant operational conditions [17]. Burton et al. and related work emphasize that uncertainty in data and models must be explicitly represented in the assurance case and mitigated via redundancy, monitoring, and fallback behaviors [18], [19]. Together, these foundations provide concepts that are directly relevant for large language model features, but they do not yet address the specific interaction patterns and failure modes of conversational or generative systems.

### 2.2 Evaluation and safety of large language models

Large language models have motivated a parallel line of research on evaluation that goes beyond traditional accuracy metrics. Liang et al. propose HELM, a holistic evaluation framework that organizes assessment along two axes: scenarios such as question answering, summarization, or dialogue, and desiderata such as accuracy, robustness, calibration, fairness, and efficiency [4]. Their empirical study of many models shows pronounced tradeoffs: a model that performs strongly on general knowledge may still behave poorly on safety or bias metrics. Benchmarks like MMLU and TruthfulQA further reveal that even frontier models struggle on broad expert level knowledge tests and can confidently reproduce common human misconceptions [5], [6].

Hallucination, the production of fluent but unsupported or incorrect statements, has emerged as a central concern for LLM evaluation. Methods such as SelfCheckGPT attempt zero resource detection of hallucinations by probing a model with variations of its own answer and measuring self-consistency [7]. These approaches illustrate the shift from simple output comparison to more complex oracles that consider internal reasoning or agreement across samples. At the same time, foundational reports on so called foundation models highlight that LLMs concentrate both capabilities and risks, including misuse, emergent behavior, and systemic biases [8]. Evaluation must therefore include not only task performance but also behavior under adversarial or unusual inputs.

To reduce harmful or misaligned behavior, model providers have adopted alignment techniques such as reinforcement learning from human feedback and constitutional AI, where models are trained to follow natural language principles of helpful and harmless behavior [9], [10]. These methods significantly improve safety in many everyday interactions, yet subsequent work on red teaming and adversarial prompting has shown that aligned models remain vulnerable to carefully crafted prompts and jailbreak strategies [11]–[15]. Safety alignment at training time is therefore a necessary but not sufficient condition for deployment in regulated environments. External guardrails, domain specific constraints, and independent safety testing remain required to ensure that LLM based features behave acceptably when deeply embedded in socio technical systems such as clinical or financial workflows.

## 3   RISK TAXONOMY FOR LLM FEATURES IN REGULATED DOMAINS

When LLM based functionality is embedded in regulated software, failures are best understood as manifestations of specific risk categories rather than as generic defects. This section proposes a taxonomy of six interrelated categories that reflect both observed LLM behaviour and concerns raised in healthcare and AI governance literature [3]–[5], [7]–[10], [16], [21], [25], [26], [30]. The categories are not mutually exclusive, but they help structure test objectives and safety arguments in later sections.

### 3.1   Factual errors and omissions

Factual inconsistency is one of the most widely documented failure modes of LLMs. Benchmarks such as TruthfulQA and HELM show that models frequently produce answers that are fluent yet factually incorrect or unsupported, especially for knowledge intensive or specialized queries [4], [5]. Methods like SelfCheckGPT treat hallucination as a first-class evaluation target and attempt to detect it through self-consistency checks [7].

In regulated domains, factual errors and omissions can have direct safety, financial, or compliance implications. Examples include incorrect interpretation of protocol inclusion criteria, wrong visit windows, incomplete lists of required documents, or misstatements about regulatory rules. Even if the LLM is not used for clinical diagnosis, inaccurate guidance about workflows or documentation can still trigger protocol deviations, billing errors, or audit findings [21], [26].

We define this category as any instance where an LLM output contradicts authoritative sources or omits critical facts that a competent domain expert would consider necessary for the intended task. This definition ties the risk explicitly to domain knowledge and supporting evidence rather than to surface plausibility alone.

### 3.2   Harmful or out of scope advice

Safety and alignment research has shown that LLMs can generate content that is harmful, offensive, or inconsistent with provider policies unless carefully trained and constrained [8]–[10]. Even when alignment methods such as reinforcement learning from human feedback or constitutional AI reduce unsafe behaviour in general use, red teaming studies demonstrate that models can still be prompted into giving disallowed advice, including self-harm instructions, hate speech, or detailed guidance on restricted activities [11]–[15].

In regulated software, a narrower but equally important concern is out of scope advice. For example, an LLM integrated into a clinical research platform may be intended only to answer questions about system configuration or documentation, but a user might ask for diagnostic recommendations or treatment choices. Providing such content could breach regulatory boundaries, professional practice norms, or institutional policies even if the information is technically correct [16], [26].

This category therefore covers both explicitly harmful content and any advice that goes beyond the declared functional scope of the feature. From a risk management perspective, systems should be designed to refuse or redirect such queries, making this an explicit target for testing and guardrail design.

### 3.3 Privacy and security risks

LLMs operating over sensitive data pose privacy and security risks even when the surrounding infrastructure complies with standard protections. Cybersecurity analyses of LLM use in healthcare highlight threats such as inadvertent disclosure of protected health information, model inversion or extraction attacks, and prompt injection that induces the system to reveal confidential content [16]. More broadly, AI trustworthiness frameworks emphasise that confidentiality, integrity, and availability must be considered alongside functional performance [25], [26].

In practical terms, privacy and security risks manifest when an LLM reproduces identifiers from input records in contexts where they should be masked, when logs capture sensitive prompts or outputs beyond authorised retention periods, or when external model APIs are used in ways that conflict with data residency or consent requirements [20], [26]. Prompt injection and other adversarial input patterns can also cause the model to ignore system instructions and leak information meant to remain hidden.

We treat this category as encompassing any behaviour that violates organisational or legal constraints on data handling, including both accidental memorisation and deliberate exfiltration via adversarial interaction.

### 3.4 Bias and unfairness

Bias and inequity in medical AI have been extensively documented, with implications for clinical decision making and patient outcomes [30]. Governance bodies such as WHO explicitly list fairness and equity as core principles for AI in health and call for active monitoring of disparate performance across subgroups [26]. LLMs trained on large scale web data inherit and can amplify social biases present in their training corpora, which can surface in tone, content, or differing quality of advice.

For LLM features in regulated software, bias and unfairness may appear in subtle ways. For instance, an assistant that helps staff draft patient communications might use systematically different language for certain demographic groups, or a configuration advisor might provide more complete guidance for large academic centres than for small community sites based on biased examples seen during training. Even if the model does not make clinical decisions directly, these patterns can contribute to unequal experiences or outcomes.

We define this category as systematic differences in behaviour across protected or contextually salient groups that are not justified by clinically or operationally relevant factors. Addressing it requires targeted testing beyond aggregate accuracy metrics.

### 3.5 Instability under change and drift

Machine learning literature distinguishes between data drift, where the input distribution changes over time, and concept drift, where the relationship between inputs and desired outputs evolves [3], [27]. Studies of clinical prediction models show that such drift can significantly degrade performance in practice if models are not regularly recalibrated or revalidated [21], [27].

For LLM features, instability arises at several levels. Model providers periodically release new versions with different capabilities and failure modes. Prompt templates and retrieval pipelines may be updated during product evolution. The data corpus used for retrieval augmented generation can change as documentation is updated. From the perspective of a regulated product, these changes can alter behaviour without any visible code modification in the host application.

This category captures risks where the system passes tests at one point in time but later degrades due to updates or environmental changes. In regulated environments, this undermines assumptions of validated performance and may require re-assessment or even regulatory notification if behaviour shifts significantly [21], [24], [26]. A taxonomy that explicitly recognises instability as a risk encourages the design of regression tests and monitoring mechanisms that track behaviour over time rather than only at initial release.

### 3.6 Adversarial and misuse risks

Red teaming work has made it clear that LLMs are susceptible to a wide range of adversarial or misuse scenarios. Researchers have demonstrated jailbreak prompts that bypass safety instructions, universal trigger phrases that provoke policy violations, and transfer attacks that work across multiple aligned models [11]–[13]. Surveys on red teaming methods describe systematic processes for exploring, establishing, and exploiting such vulnerabilities in order to improve defences [11]–[15].

In regulated software, adversarial behaviour may come from external attackers but also from insiders experimenting with the system, curious users, or even benign queries that inadvertently mimic attack patterns. Prompt injection can cause an LLM to ignore system constraints and act on untrusted instructions embedded in retrieved content. Combined with the other risk categories, adversarial prompts can turn otherwise unlikely failure modes into probable events, for example by forcing the model to reveal private information or to give out of scope medical advice [16].

We define this category as behaviour that results from intentional or unintentional exploitation of model or system weaknesses through crafted inputs or interactions. Recognising adversarial and misuse risks as a distinct category helps justify dedicated stress testing and red teaming efforts rather than assuming that normal functional tests will suffice.

### 3.7 Summary of risk categories

The six categories described above provide a structured lens for reasoning about LLM based features in regulated environments. In practice, individual incidents often span several categories, such as a jailbreak prompt that elicits both a factual error and a privacy breach. For testing and assurance, however, treating each category as a separate test objective allows teams to design focused artefacts such as golden query sets for factual accuracy, policy violation suites for harmful advice, privacy leakage tests, bias probes, regression dashboards for stability, and adversarial prompt corpora. The next section builds on this taxonomy to derive a concrete test strategy and architecture.

## 4 TEST STRATEGY AND ARCHITECTURE FOR LLM FEATURES

The risk taxonomy in Section 3 provides a vocabulary for describing failures, but engineering teams need concrete test artefacts and architectural patterns that operationalize these risks. In this section we propose a test strategy organized along two dimensions. The first is an architectural view that separates guardrails, LLM orchestration, and the surrounding application. The second is a mapping from each risk category to specific test types that can be automated and integrated into the software lifecycle. The design is informed by work on machine learning testing and safety assurance [1]–[3], [17]–[19] as well as LLM evaluation and red teaming [4], [5], [7], [11]–[15].

## 4.1 Architectural layers for testing LLM features

We consider LLM based functionality as a subsystem composed of three interconnected layers. Each layer has distinct responsibilities and corresponding test objectives.

1. **Guardrail and policy layer**: This layer enforces organizational and regulatory constraints on what the system may accept and produce. It typically includes input validation, content filters, allow and block lists, and explicit policy checks for privacy, safety, and scope. In regulated domains this layer is the primary line of defense against harmful or out of scope advice, privacy leaks, and obvious policy violations [16], [20], [25], [26].

2. **Prompt orchestration and retrieval layer**: This layer constructs prompts, manages retrieval augmented generation, and post processes model outputs. It determines which context is provided to the LLM, how instructions are phrased, and how responses are transformed into structured outputs. Errors here often manifest as factual inconsistencies, omissions, or unstable behavior across updates, since small changes in templates or retrieval logic can significantly affect outputs [4], [5], [7], [21].

3. **System and user experience layer**: This layer integrates the LLM functionality into the broader application, controls how and when the feature is invoked, and determines what the user sees. It includes user interface elements, explanation mechanisms, logging, and monitoring. Failures here can expose users to misinterpreted outputs, hide uncertainty, or fail to record evidence needed for audits and incident investigation [3], [20], [21], [24], [26].

From a safety assurance perspective, this decomposition aligns with the idea that safety arguments should cover both component internals and system context [17]–[19], [25]. The guardrail layer corresponds to explicit safety constraints, the orchestration layer to model centered validation, and the system layer to human factors and real world integration. Tests should be scoped to the layer most directly responsible for a given risk, while still recognizing cross layer interactions.

## 4.2 Test types per risk category

Using the six risk categories from Section 3, we now outline test types that can be attached to each category and mapped to the architectural layers above. In practice, organizations can start with a subset that addresses their highest risks, then expand coverage over time.

### 4.2.1 Factual accuracy and completeness tests

To address factual errors and omissions, we propose golden-set-based tests and retrieval consistency tests.

Golden sets comprise representative queries paired with reference answers or required facts curated by domain experts or derived from authoritative documents [3], [4], [5], [28]. For each query, the test harness checks that the LLM output satisfies constraints such as inclusion of specific values, absence of contradictions, and coverage of mandatory elements. For example, configuration queries about visit windows can be tested to ensure that the correct numerical ranges and conditions appear in the response.

Retrieval consistency tests target systems that use retrieval augmented generation. For a given query and document set, the test verifies that the correct documents are retrieved and cited, and that the generated answer is consistent with those sources [4], [7]. Discrepancies where the model introduces content not grounded in retrieved documents are treated as hallucination candidates.

These tests primarily exercise the orchestration and retrieval layer but rely on the system layer for appropriate logging and on the guardrail layer to block clearly unsupported claims. They can be run offline on large batches of recorded or synthetic queries and integrated into regression testing whenever prompts, retrieval logic, or model versions change.

### 4.2.2 Harmful and out of scope advice tests

For harmful or out of scope advice, the main test artefact is a policy violation suite. This consists of prompts that intentionally probe boundaries the system must not cross, such as requests for diagnosis, treatment recommendations, or prohibited activities in the clinical or organizational context [9]–[12], [15], [16], [26].

Expected behavior for each prompt is defined as refusal, safe redirection, or a policy compliant alternative. Automated checks can look for the presence of refusal patterns, disclaimers, and absence of disallowed keywords or phrases in outputs. Guardrail components such as content classifiers and rule-based filters can be unit tested with these prompts, while end to end tests validate that the combined guardrail and LLM output remains within scope.

Red teaming literature suggests that such suites should be periodically expanded with new prompts discovered through manual or automated adversarial exploration, since static sets tend to become stale as models and usage evolve [11]–[15]. In regulated products, these suites also serve as evidence that known high risk behaviors have been explicitly tested and mitigated.

### 4.2.3 Privacy and security tests

Privacy and security tests aim to detect leakage of sensitive information, improper handling of identifiers, and susceptibility to prompt injection. Inspired by security testing and healthcare specific analyses of LLM threats [16], [20], [25], [26], we recommend the following elements.

First, synthetic sensitive data tests. Test inputs and retrieval corpora are seeded with synthetic identifiers, such as fictitious names, addresses, or record numbers that follow realistic formats. Prompts then attempt to elicit these values directly or indirectly, for example by asking for summaries, lists, or verbatim reproduction. Automated checks flag any occurrence of synthetic identifiers in contexts where they should not appear, indicating potential memorization or inappropriate exposure.

Second, prompt injection tests. These tests construct inputs where untrusted text includes instructions to ignore previous policies, reveal hidden information, or execute unwanted actions, similar to those described in red teaming work [11]–[13]. The test passes only if the system honors system level constraints and ignores the injected instructions. This often involves unit tests on prompt sanitization and end to end tests on the combined pipeline.

Third, logging and audit tests. Here the focus shifts to the system layer. Test cases verify that logs and telemetry appropriately mask or exclude sensitive content, and that audit records capture enough detail about prompts, model versions, and outputs to support incident investigation without exposing protected data [20], [24], [26].

### 4.2.4 Bias and fairness tests

Bias and unfairness require tests that compare behavior across subgroups rather than evaluating individual outputs in isolation. Following recommendations from bias studies in medical AI and governance documents [26], [30], we propose paired prompt testing and subgroup performance analysis.

Paired prompts are constructed by varying demographic attributes or other protected characteristics while keeping the substantive query constant. For example, prompts may describe otherwise identical patient scenarios differing only in age, gender, or ethnicity. Outputs are then analyzed for differences in tone, content, or recommendation completeness. Automated metrics can count occurrences of certain adjectives or measure length and structure, while domain experts perform qualitative review of a sample for more subtle patterns [30].

When the LLM produces structured outputs, such as suggestions for follow up steps or document lists, subgroup performance metrics can be defined analogously to classification performance metrics. For instance, one can compute the rate at which critical elements are included for each subgroup. Significant discrepancies trigger investigation and potentially further data curation or prompt adjustments.

These tests typically operate at the orchestration and system layers, but their design and interpretation are deeply tied to organizational commitments about equity and non-discrimination.

#### 4.2.5 Stability and regression tests

Instability under change motivates regression tests that track behavior over time. In line with continuous validation ideas from AI system assurance and clinical model monitoring [3], [21], [27], we propose maintaining a frozen regression suite and performing periodic differential evaluation.

The regression suite combines elements from factual, policy, privacy, and bias tests into a compact set of high value cases that represent core workflows and previously observed failures. Whenever any of the following change, the suite is executed automatically:

- LLM provider or model version.
- Prompt templates or retrieval ranking algorithms.
- Major updates to the underlying document corpus.

For each new configuration, outputs are compared to prior baselines using the same oracles as in the original tests. The goal is not to enforce identical text but to detect breaches of constraints such as newly introduced factual errors, weakened refusals, increased leakage of sensitive patterns, or regression in bias metrics. In regulated settings, this process also supports documentation of how model updates are evaluated and either accepted or rolled back [20], [21], [24], [26].

#### 4.2.6 Adversarial and red team tests

Finally, adversarial and misuse risks call for dedicated red team activities that complement the more structured test suites above. Building on work that uses LLMs themselves to generate challenging prompts and explores universal jailbreak attacks [11]–[15], organizations can adopt a two stage approach.

In the exploration stage, testers or automated adversaries search for prompts and interaction patterns that cause the system to violate safety, privacy, or scope constraints. Successful attacks are triaged, and a subset is translated into reproducible test cases. In the consolidation stage, these cases are added to the policy, privacy, or factual suites as appropriate, effectively converting discovered vulnerabilities into regression tests.

This approach turns red teaming from an ad hoc activity into a continuous source of new test cases. In regulated products, records of red team exercises and subsequent hardening can be included in safety arguments and risk registers [17], [19], [20], [25].

#### 4.2.7 Summary mapping of risks to test types

Table 1 summarises how the six risk categories from Section 3 map to concrete test types and architectural layers. It also indicates which tests are suitable for automation and which typically require expert review.

Table 1. Mapping of risk categories to test types and architectural layers

| Risk category | Example failures in Knowledgebase assistant | Primary test types | Main architectural layers | Automation potential |
|---|---|---|---|---|
| Factual errors and omissions | Mixing legacy and current workflows, missing required navigation steps, incorrect parameter ranges | Golden query and answer sets; retrieval consistency tests; source alignment checks | Orchestration and retrieval; system logging | High for pattern and constraint checks; expert review for new domains |
| Harmful or out of scope advice | Clinical commentary when only product guidance is allowed, suggestions that undermine protocol or policy | Policy violation suites; refusal pattern checks; safety and scope guardrail tests | Guardrail and policy; end to end system | High for refusal pattern and keyword checks; medium for nuanced content |
| Privacy and security risks | Echoing synthetic identifiers, reproducing internal examples verbatim, prompt injection overriding system instructions | Synthetic sensitive data leakage tests; prompt injection tests; logging and audit tests | Guardrail; orchestration sanitisation; system logging | High for identifier patterns and injection behaviour; expert review for edge cases |
| Bias and unfairness | Richer guidance for large academic centres than for small community sites, uneven tone across site types or regions | Paired prompt tests; subgroup performance comparison; language and tone analysis | Orchestration; system UX and analytics | Medium for automatic metrics; requires expert review for interpretation |

| Instability under change and drift | Degraded refusals and incomplete answers after LLM version swap or corpus update | Frozen regression suite across versions; differential comparison; drift and trend monitoring | Orchestration; system monitoring and deployment | High for automated regression and alerting; expert review for acceptance decisions |
|---|---|---|---|---|
| Adversarial and misuse risks | Jailbreak prompts eliciting implementation speculation, instructions that circumvent guardrails, injected commands in logs | Red team exploration; adversarial prompt corpora; conversion of attacks into regression tests | Guardrail; orchestration; system access control | Medium automation for replay suites; exploration remains partly manual or semi automated |

## 4.3 Automation and integration into the software lifecycle

To be practical, the test strategy must fit into existing development and validation workflows rather than remaining a one-time exercise. Machine learning testing surveys emphasize the importance of integrating tests at multiple levels of the stack and throughout the lifecycle [1]–[3]. For LLM features, we suggest the following integration points.

At development time, unit tests cover guardrail functions, prompt construction utilities, retrieval and ranking components, and post processing logic. These tests resemble traditional software tests and can be run on every code change.

At integration level, batch evaluation jobs run factual, policy, privacy, and bias suites against development and staging deployments. These jobs can be triggered by changes in configuration or dependencies, such as a new model version, and their reports act as gates for promotion to production.

At system level, pre-release validation can incorporate scenario-based tests and, where appropriate, limited pilots or shadow deployments that compare LLM assisted workflows with baseline processes, following ideas from clinical validation and dynamic deployment of medical AI [3], [21], [27].

Post deployment, monitoring pipelines track key indicators such as user feedback, refusal rates, distribution of query types, and drift signals, and periodically rerun selected tests on live configurations. Alerts are raised when metrics cross thresholds or when tests that previously passed begin to fail. These mechanisms support continuous assurance in line with emerging expectations from regulators and governance frameworks [20], [21], [24]–[26].

Taken together, the layered architecture and mapped test types provide a blueprint for systematic testing of LLM features in regulated software. The next section connects these ideas more explicitly to validation and governance practices in healthcare and related domains.

# 5 VALIDATION IN REGULATED AND SAFETY CRITICAL ENVIRONMENTS

The test strategy in Section 4 outlines how to structure and automate checks for LLM based features at subsystem level. In regulated environments, however, testing is embedded in a broader validation and assurance process that spans the entire lifecycle of the product. This section connects the proposed strategy to existing practices and expectations in healthcare and related safety critical sectors, with an emphasis on how test artefacts contribute to safety arguments, clinical or operational evidence, and governance.

## 5.1 Lessons from medical device AI validation

Over the past decade, regulatory and industry communities have accumulated experience with validation of AI enabled medical devices, particularly for image analysis and structured prediction tasks. Surveys of the regulatory landscape show a growing number of AI based devices cleared by agencies such as the FDA and notified under European frameworks, often classified as software as a medical device [22]–[24]. Higgins and Johner highlight that across pharmaceuticals, medical devices, and in vitro diagnostics, AI containing products are expected to follow discipline specific quality systems while also addressing novel questions about training data, performance drift, and human factors [20].

Common themes in this literature include the need for validation datasets that reflect intended use, careful separation of training and test data, and demonstration of performance on clinically relevant endpoints rather than proxy metrics [20], [21], [28]. Homeyer et al. provide detailed recommendations for compiling test datasets in pathology, emphasizing coverage of case diversity, rarity, and technical variability so that performance estimates are meaningful when the tool is deployed [28]. Myllyaho et al. and Rosenthal et al. further point out that validation rarely ends at offline testing; simulation environments, prospective studies, and expert review of tool outputs in context play important roles in building confidence for high consequence use [3], [21].

These patterns suggest several implications for LLM features. First, golden sets and policy suites should be constructed to reflect the tasks and populations that the feature will actually support, not only synthetic or simplified examples. Second, performance should be assessed in terms that matter to the regulated domain, such as reduction in protocol deviations or error rates in documentation, rather than solely in generic language metrics. Third, where LLM outputs influence clinical workflows, it may be appropriate to include them in broader validation activities such as simulated user studies or limited real world pilots, in line with how other AI tools are assessed [20], [21].

## 5.2 Towards trial like evaluation for LLM features

Traditional clinical AI validation often assumes a fixed model evaluated once in a prospective trial. Rosenthal et al. argue that this assumption is increasingly unrealistic as AI systems evolve after deployment and are adapted to local contexts [21]. They propose dynamic deployment paradigms in which models can be updated under trial governance, with continuous monitoring and adaptation rather than a single frozen evaluation. Similar concerns apply to LLM features, where model providers release new versions, organizations adjust prompts, and document corpora change over time.

For LLM based features embedded into clinical or research platforms, full randomized trials may not always be feasible, yet some of the same principles can be applied at smaller scale. For high-risk use cases, organizations can use shadow deployments in which the LLM runs alongside existing workflows and its outputs are recorded but not acted upon initially. This allows assessment of factual accuracy, safety, and impact on workflow without patient risk, analogous to shadow mode validation in other AI systems [3], [21], [27].

Where LLM features directly influence user actions, such as suggesting configuration changes or summarizing protocol text, limited rollouts with careful monitoring and predefined stopping criteria can provide evidence of benefit and safety. Metrics might include error rates, time to complete tasks, user satisfaction, and incidence of safety incidents or near misses. These evaluations complement the pre release test suites by measuring performance under real use conditions, including unanticipated query types and user behaviors. The regression and monitoring mechanisms described in Section 4.2.5 and 4.3 then support ongoing surveillance, making validation a continuous process rather than a one time gate [3], [21], [27].

### 5.3 Governance, documentation, and auditability

Regulatory and governance frameworks for AI in health stress not only technical performance but also transparency, accountability, and risk management. The WHO guidance on ethics and governance of AI for health calls for clear documentation of data sources, model behavior, evaluation methods, and limitations, alongside explicit consideration of bias, equity, and human oversight [26]. Broader AI trustworthiness frameworks similarly emphasize risk identification, control measures, and monitoring, supported by artefacts that can be examined by internal or external auditors [25].

Within this context, the test strategy proposed in Section 4 can be seen as a mechanism for generating and maintaining a structured set of assurance artefacts. Golden sets, policy suites, privacy and bias test results, regression reports, and records of red team exercises collectively provide evidence that concrete risks have been identified, probed, and mitigated. When associated with specific LLM configurations, prompt templates, and deployment environments, these artefacts also support traceability over time, which is critical when explaining how a given feature was validated at a particular point [20], [21], [24]–[26].

From a governance perspective, two additional aspects are important. First, responsibilities for designing, executing, and reviewing tests should be clearly assigned, for example distinguishing between development teams, independent validation or QA groups, and domain experts. This aligns with model risk management practices in finance and safety assurance approaches such as AMLAS, which advocate separation of roles and independent challenge of safety arguments [17], [19], [20], [25]. Second, organizations should define thresholds and decision rules that connect test results to actions, such as blocking release, requiring further review, or initiating incident response. Without such policies, even well-designed tests may fail to influence practice.

Overall, aligning LLM testing with established validation and governance practices in regulated domains helps ensure that these features are treated as first class, safety relevant components rather than experimental add-ons. In the next section we discuss practical implications of adopting this framework, including tradeoffs and limitations.

## 6 DISCUSSION AND IMPLICATIONS FOR PRACTICE

The proposed framework translates high level concerns about LLM behaviour into concrete test objectives and artefacts for regulated software. In this section we reflect on practical implications, trade offs, and limitations of adopting such a framework in real organisations.

### 6.1 Balancing test depth with delivery cadence

One tension for product teams is the balance between thorough testing and the desire to deliver new LLM features and improvements rapidly. Machine learning testing surveys already highlight the cost of extensive data collection, expert labelling, and test maintenance [1]–[3]. For LLM based features, the need for curated golden sets, policy suites, and bias probes adds additional overhead.

A pragmatic approach is to prioritize test development based on risk. High consequence behaviors, such as guidance that influences clinical workflows or access to sensitive data, should receive extensive factual, policy, and privacy testing. Lower risk interactions, such as convenience features that merely rephrase documentation, may initially be covered by lighter test suites and monitoring. Over time, organizations can expand coverage as resources allow.

Change management can also be structured by defining tiers of change with corresponding test requirements, a pattern seen in medical AI validation and model risk management [20], [21], [24], [25]. For example, switching to a new LLM provider or altering the functional scope of the feature might trigger full regression and red team testing, while minor prompt tweaks might require only focused subsets of the suite and monitoring of key indicators.

## 6.2 Integrating domain expertise into testing

Many of the proposed tests, especially for factual accuracy, bias, and out of scope advice, rely on domain expertise. This mirrors observations from AI validation literature, where expert review and clinical judgement are central to assessing performance and safety [3], [20], [21], [28], [30]. For LLM features, domain experts are needed to curate golden answers, define prohibited topics, interpret borderline outputs, and assess fairness trade offs.

This dependence on experts can become a bottleneck, particularly in organisations with limited specialist availability. One mitigation is to design test artefacts that maximise reuse of expert effort. For instance, a single expert review session can produce multiple test cases and reference answers that become part of the long lived golden set. Similarly, red team sessions with domain experts can generate prompt patterns and failure examples that later feed into automated suites.

Another strategy is to combine expert input with automated tools. Techniques such as SelfCheckGPT and embedding based similarity can flag outputs that deviate from known patterns for expert review [4], [7]. This allows experts to focus on the most informative cases rather than manually checking every test.

## 6.3 Limitations of current techniques and residual risk

Despite expanding test coverage, some residual risks remain difficult to eliminate. Adversarial and misuse risks are particularly challenging, as new jailbreak prompts and attack strategies continue to emerge [11]–[15]. Red teaming can identify and mitigate many vulnerabilities, but it cannot guarantee complete coverage. Similarly, bias and fairness tests can reveal disparities across selected groups, yet may miss issues in untested contexts or interactions [26], [30].

Hallucination detection methods are also imperfect. Approaches that rely on self-consistency or external retrieval can reduce obvious factual errors, but they may fail when sources are ambiguous, incomplete, or themselves biased [4], [5], [7], [8]. This suggests that for high consequence decisions, LLM outputs should be treated as advisory and subject to human verification, in line with recommendations from governance bodies [21], [25], [26].

From a regulatory standpoint, it is unlikely that any test strategy will remove all uncertainty about LLM behavior. Instead, the goal is to reduce risk to an acceptable level, document the residual uncertainties, and put in place processes for monitoring, incident response, and improvement. Safety assurance frameworks for machine learning already adopt this stance, focusing on structured arguments and evidence rather than absolute guarantees [17]–[19], [25]. The proposed test framework aims to complement such assurance cases with concrete, repeatable checks for LLM features.

### 6.4 Generalizability beyond healthcare

Although the paper focuses on healthcare and clinical research as exemplar regulated domains, the underlying risk categories and test patterns are relevant to other sectors. Financial services, for example, face similar concerns about factual errors in regulatory guidance, privacy breaches in customer interactions, and bias in advisory outputs. Model risk management guidelines in banking already require independent validation, stress testing, and documentation that are conceptually aligned with the proposed approach [20], [24], [25].

Safety critical industries such as automotive and aviation may also adopt LLM components for documentation, maintenance support, or operator assistance. In these settings, the combination of factual, policy, and adversarial tests, along with integration into safety cases, can support arguments that the LLM components do not undermine overall system safety [17]–[19]. Adapting the framework to such domains mainly involves tailoring risk scenarios and test artefacts to local regulations and operational practices.

### 6.5 Threats to validity

As with any conceptual framework, several threats to validity should be noted. First, the taxonomy of risks is derived from existing literature and emerging practice, and may not fully capture all relevant failure modes, especially as LLM capabilities and usage patterns evolve [4], [8], [11]–[15]. Second, the described test types assume access to logs, configuration control, and some level of influence over model selection and prompts, which may not hold for all deployments, particularly when using externally hosted models with limited transparency.

Third, while the framework draws on principles from medical AI validation and safety assurance, it has not yet been empirically evaluated across multiple organizations or use cases. Comparative studies that assess its impact on defect rates, incident occurrence, and regulatory reviews would strengthen confidence in its effectiveness. Finally, the framework does not prescribe specific metrics or thresholds, leaving room for inconsistent interpretations. Future work could refine quantitative indicators and decision rules for different risk levels and application types.

# 7 CONCLUSION

Large language models are increasingly embedded into software products used in regulated and safety critical domains. Their strengths in natural language understanding and generation come with distinctive risks, including hallucinated content, harmful or out of scope advice, privacy and security issues, bias, instability under change, and vulnerability to adversarial prompts. Existing machine learning testing and AI validation literature provides valuable foundations but does not fully address the interactive, generative, and rapidly evolving nature of LLM features in these settings.

This paper has proposed a risk-based taxonomy tailored to LLM features in regulated software, along with a corresponding test strategy and architectural decomposition. By mapping each risk category to concrete test types and associating them with guardrail, orchestration, and system layers, the framework offers a blueprint for designing golden sets, policy suites, privacy and bias probes, regression suites, and red team exercises that can be integrated into existing development and validation workflows.

Connecting this strategy to practices in medical device AI validation and broader governance frameworks underscores that testing is only one element of assurance. Test artefacts must be complemented by clear roles, decision rules, monitoring, and documentation to satisfy regulatory and organizational expectations. Nonetheless, systematic testing is a necessary foundation for any credible safety argument about LLM based features.

Future work includes empirical evaluation of the framework in real deployments, development of shared benchmarks and test corpora for regulated use cases, and exploration of how automated tools can support test maintenance as models, prompts, and regulations evolve. As LLMs continue to be adopted in high stakes contexts, collaborative efforts between software engineers, domain experts, and regulators will be essential to ensure that their benefits are realized without compromising safety, privacy, or fairness.